\documentstyle[12pt]{article}
\input{epsfig.sty}
\newcommand{\beq}{\begin{eqnarray}}
\newcommand{\eeq}{\end{eqnarray}}
\begin{document}
\title{ $\Psi$ and $\Upsilon$ Production in Proton-Proton Collisions at 
E=13 TeV}
\author{Leonard S. Kisslinger$^{1}$\\
Department of Physics, Carnegie Mellon University, Pittsburgh PA 15213 USA.\\
Debasish Das$^{2,3}$\\
High Energy Nuclear and Particle Physics Division,\\
 Saha Institute of Nuclear 
Physics, 1/AF, Bidhan Nagar, Kolkata 700064, INDIA.}
\date{}
\maketitle

\begin{abstract}
  This article is an extension of our recent studies of  $\Psi$ and
$\Upsilon$ production cross sections in proton-proton collisions at the LHC 
with E=$\sqrt{s}$=8.0 TeV to E=13 TeV 
\end{abstract}

\noindent
1) kissling$@$andrew.cmu.edu \hspace{1cm} 2)dev.deba$@$gmail.com; 
3) debasish.das@saha.ac.in
\vspace{2mm}

\vspace{2mm}

\noindent
PACS Indices:12.38.Aw,13.60.Le,14.40.Lb,14.40Nd

\section{Differential Rapidity Cross Sections for $J/\Psi,\Psi(2S)$ 
Production Via p-p Collisions at E= 13 TeV}

  The present work is an extension of our previous 
study of $\Psi$ and $\Upsilon$ production via p-p Collisions at $\sqrt{s}$
=E= 8.0 TeV\cite{kd14} to 13 TeV. We use the theory described in detail in 
Ref\cite{klm11} based on the octet model\cite{cl96,bc96,fl96} for p-p 
production of heavy quark states. 

The differential rapidity cross section for the  $J/\Psi(1S)$ 
meson production for the helicity $\lambda=0$  is given by~\cite{klm11}.

\beq
\label{dsig}
      \frac{d \sigma_{pp\rightarrow \Psi(\lambda=0)}}{dy} &=& 
     A_\Phi \frac{1}{x(y)} f_g(x(y),2m)f_g(a/x(y),2m) \frac{dx}{dy} \; ,
\eeq 
where $y$=rapidity, $a= 4m^2/s$, $s=E^2$, $A_\Phi=\frac{5 \pi^3 \alpha_s^2}
{288 m^3 s}\;<O_8^\Phi(^1S_0)>$,  with $\alpha_s$=.118, $\;\;\;\;$
 $<O_8^\Phi(^1S_0)>$=.0087$GeV^3$, and $f_g$ is the gluonic distribution 
function.  In the present work $E=\sqrt{s}=13.0$ TeV and for $J/\Psi,\Psi(2S)$
production $m\simeq$ 1.5 GeV.

    The rapidity variable, $y$, used for differential cross sections, is
\beq
\label{rapidity}
      y(x) &=& \frac{1}{2} ln (\frac{E + p_z}{E-p_z}) \nonumber \\
        E &=& \sqrt{s} = \sqrt{m^2 + p_z^2}  \nonumber \\
             p_z &=& \frac{\sqrt{s}}{2} (x-\frac{a}{x})  \; .
\eeq
\newpage

The functions of rapidity $x(y), \frac{dx}{dy}$ are
\beq
\label{4}
   x(y) &=& 0.5 \left[\frac{m}{E}(\exp{y}-\exp{(-y)})+\sqrt{(\frac{m}{E}
(\exp{y}-\exp{(-y)}))^2 +4a}\right] \nonumber \\
  \frac{d x(y)}{d y} &=&\frac{m}{2E}(\exp{y}+\exp{(-y)})\left[1. + 
\frac{\frac{m}{E}(\exp{y}-\exp{(-y)})}{\sqrt{(\frac{m}{E} 
(\exp{y}-\exp{(-y)}))^2 +4a}}\right] \; .
\eeq

The gluonic distribution $f_g(x(y),2m)$ for the range of $x$ needed for 
$E=13.0$ TeV is\cite{klm11}
\beq
\label{fgupsilon}
         f_g(x(y))&=& 275.14 - 6167.6*x + 36871.3*x^2 \; .
\eeq

Using the method of QCD sum rules it was shown\cite{lsk09} that
the $\Psi(2S)$ is a 50\%-50\% mixture (with approximately a 
10\% uncertainty) of standard quarkonium and hybrid quarkonium states.
This solved the problem of the branching ratios of hadronic decays of
the $\Psi(2S)$ to the $J/\Psi$. Thus the  $\Psi(2S)$ state is 
\beq
    |\Psi(2S)>&=& -0.7 |c\bar{c}(2S)>+\sqrt{1-0.5}|c\bar{c}g(2S)> \; ,
\eeq
while the $J/\Psi$ state is essentially a standard $c \bar{c}$ state
\beq
        |J/\Psi(1S)>&=& |c\bar{c}(1S)> \; ,
\eeq 
with $c$ a charm quark.  Also in Ref\cite{lsk09} it was shown that the 
$\Upsilon(3S)$ state is approximately 50\%-50\% mixture of a standard
and hybrid bottomonium state, 
\beq
    |\Upsilon(3S)>&=& -0.7 |b\bar{b}(3S)>+\sqrt{1-0.5}|b\bar{b}g(3S)> \; ,
\eeq
with $b$ a bottom quark, which we refer to as a mixed heavy quark hybrid state,
while the $\Upsilon(1S)$ and $\Upsilon(2S)$ are standard $b\bar{b}$ states.
This solved the problem of $\sigma$ decays of $\Upsilon$ states\cite{lsk09}. 

For $J/\Psi(1S)$ production $A_\Phi=A_{\Psi(1S)}\simeq 1.85 x 10^{-7}$ nb.
For $\Psi(2S)$ production with the standard model $A_{\Psi(2S)}\simeq
0.039 A_{\Psi(1S)}$, while with the mixed hybrid theory $A_{\Psi(2S)}\simeq
0.122 A_{\Psi(1S)}$. 

With the parameters given above for $J/\Psi(1S)$ and $\Psi(2S)$  production, 
from Eq(\ref{dsig}),  $\frac{d \sigma_{pp\rightarrow \Psi(1S)}}{dy}$ and 
$\frac{d \sigma_{pp\rightarrow \Psi(2S)}}{dy}$, are shown in Figure 1. Note 
that $\frac{d \sigma_{pp\rightarrow \Psi(2S-standard)}}{dy}$ is much smaller than
$\frac{d \sigma_{pp\rightarrow \Psi(2S-hybrid)}}{dy}$, which is important for
studies of the possible production of the Quark Gluon Plasma (QGP)in
Relativistic High Energy Collisions (RHIC). See, e.g., Ref\cite{klm14}.

\section{Differential Rapidity Cross Sections for $\Upsilon(1S),
\Upsilon(2S),\Upsilon(3S)$ Production Via p-p Collisions at E= 13
 TeV}

The differential rapidity cross sections for $\Upsilon$ production are given
by Eq(\ref{dsig}) with  $m\simeq$ 5 GeV. $A_{\Upsilon(1S)}\simeq 5.0x10^{-9}$, 
$A_{\Upsilon(2S)}\simeq 0.039xA_{\Upsilon(1S)}$,  $A_{\Upsilon(3S)}\simeq 
0.0064xA_{\Upsilon(1S)}$ for the standard model, and $A_{\Upsilon(3S)}\simeq 
0.012xA_{\Upsilon(1S)}$ for the
mixed hybrid theory. $\frac{d \sigma_{pp\rightarrow \Upsilon(nS)}}{dy}$ are shown
in Figure 2. Note that $\frac{d \sigma_{pp\rightarrow \Upsilon(3S-standard)}}{dy}$ is 
much smaller than $\frac{d \sigma_{pp\rightarrow \Upsilon(3S-hybrid)}}{dy}$, which 
also is used for studies of the possible production of the QGP via RHIC.

\newpage

The differential rapidity cross sections for $J/\Psi(1S)$ and $\Psi(2S)$  
production for the standard model and the mixed hybrid theory are shown in 
Figure 1.
\vspace{11cm} 

\begin{figure}[ht]
\begin{center}
\epsfig{file=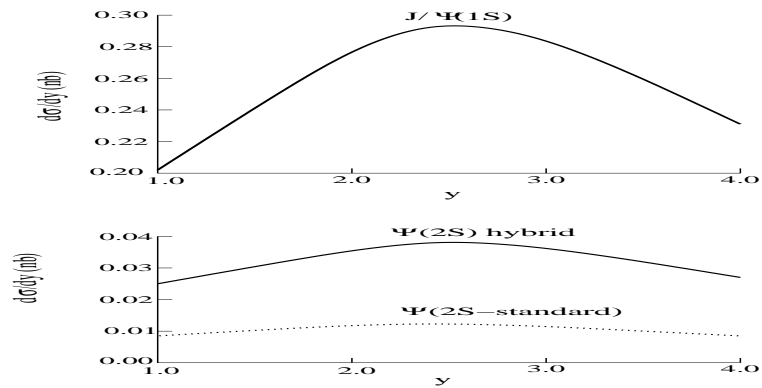,height=5cm,width=10cm}
\end{center}
\caption{d$\sigma$/dy for p-p collisions at $\sqrt{s}$ = 13.0 TeV 
producing $J/\Psi(1S)$; and $\Psi(2S)$ for the standard model (dashed curve) 
and the mixed hybrid theory.}
\label{Figure 1}
\end{figure}   

\newpage

  Figure 2,  differential rapidity cross sections for Upsilon production
\vspace{10cm}

\begin{figure}[ht]
\begin{center}
\epsfig{file=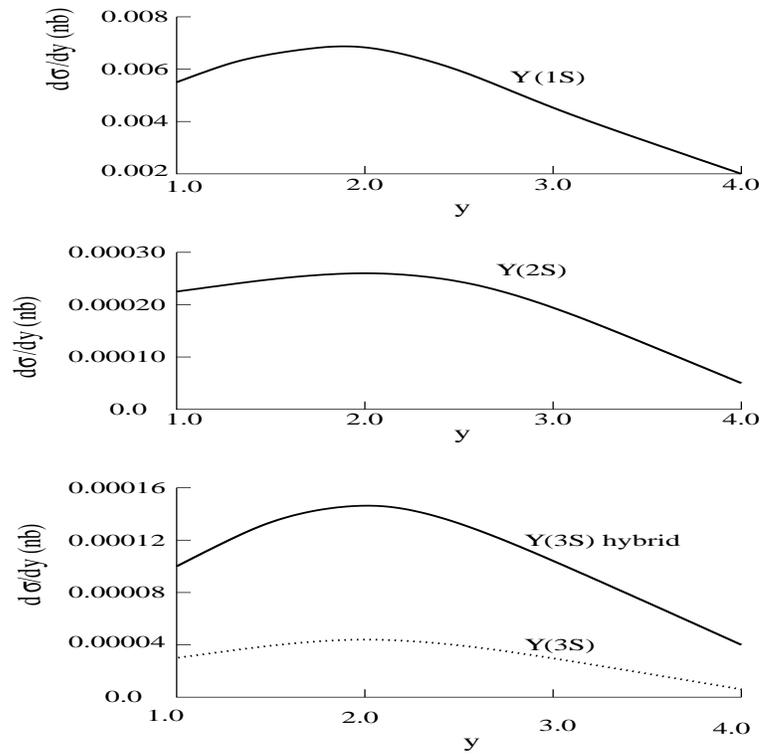,height=10cm,width=10cm}
\end{center}
\caption{d$\sigma$/dy for p-p collisions at $\sqrt{s}$ = 13.0 TeV 
producing $\Upsilon(1S)$, $\Upsilon(2S)$, and $\Upsilon(3S)$ for the standard 
model (dashed curve) and the mixed hybrid theory.}
\label{Figure 2}
\end{figure}

\newpage
  \section{Ratio of  $\Psi(2S)$ to $J/\Psi(1S)$ and $\Upsilon(2S),
\Upsilon(3S)$  to $\Upsilon(1S)$ cross sections}

Since there is uncertainty in the magnitude of the  cross sections, shown
in Figures 1 and 2, an essential test of our mixed heavy hybrid theory for 
heavy quark production are the ratios of cross section for the production of 
heavy quark states.

  As discussed in earlier publications\cite{kd13,kd14,klm11} the estimated
$\Psi(2S)$ to $J/\Psi(1S)$ ratios for the standard nodel and the mixed
hybrid theory for p-p production are
\beq
\label{ppratio}
    \sigma(\Psi(2S))/\sigma(J/\Psi(1S))|_{standard} &\simeq& 0.27 
\nonumber \\
    \sigma(\Psi(2S))/\sigma(J/\Psi(1S))|_{hybrid} &\simeq& 0.67\pm 0.07 \; .
\eeq

The ratio for the mixed hybrid theory shown in Eq(\ref{ppratio}) is 
consistent with the PHENIX experimental result for E=200 GeV\cite{phenix},
while the ratio for the standard model is not consistent with experiment.
This ratio should be approximately independent of energy\cite{klm11}.  
Presumably, this will be tested by the production of $J/\Psi(1S)$ and 
$\Psi(2S)$ states via 13 TeV p-p collisions in the near 
future \cite{pmrc14,gregl13}.

The estimated $\Upsilon(2S),\Upsilon(3S)$ to $\Upsilon(1S)$ ratios,  
as discussed in Refs\cite{kd13,klm11} are

\beq
\label{lsk11}
     \Upsilon(2S)/\Upsilon(1S)|_{standard} &\simeq& 
\Upsilon(2S)/\Upsilon(1S)|_{hybrid} \simeq 0.27 \nonumber \\
     \Upsilon(3S)/\Upsilon(1S)|_{standard} &\simeq& .04 \nonumber \\
     \Upsilon(3S)/\Upsilon(1S)|_{hybrid} &\simeq& 0.14-0.22 \; .
\eeq
These ratios have been determined in recent LHCb experiments at 
8 TeV\cite{LHCb13} and 2.76 TeV \cite{LHCb14}. These LHCb measurements show 
that for the $\Upsilon(2S)/\Upsilon(1S)$ ratio the standard model, which 
is the same in the mixed hybrid theory, is correct, while the experimental 
$\Upsilon(3S)/\Upsilon(1S)$ ratio is consistent with the mixed hybrid theory
whereas the standard model prediction is much too small. Recently the CMS 
Collaboration has measured the differential cross sections for 
$\Upsilon(1S),\Upsilon(2S),\Upsilon(3S)$\cite{CMS1-2015}
and $J/\Psi(1S),\Psi(2S)$\cite{CMS2-2015} states produced via p-p collisions
at 7 TeV, from which ratios similar to those in Eqs(\ref{ppratio},\ref{lsk11})
can be estimated.

\section{Conclusions}

  Our results for the rapidity dependence of
d$\sigma$/dy, shown in the figures, and the ratio of cross sections should be 
useful for experimentalists studying heavy quark production in p-p collisions
at 13 TeV at the LHC.  It is also a further test of the validity of the mixed 
heavy quark hybrid theory, for which at lower energy p-p collisions the 
ratios of $\sigma(\Psi(2S))/\sigma(J/\Psi(1S))$ and $\sigma(\Upsilon(3S))/
\sigma(\Upsilon(1S))$ have been shown to be in 
agreement, within errors, with the mixed hybrid theory, but not the standard 
quark-antiquark model . This is very important since we are using the mixed 
hybrid heavy quark theory to test the creation of the Quark Gluon Plasma 
via Relativistic Heavy Ion Collision experiments.
\vspace{5mm}

\Large{{\bf Acknowledgements}}

\normalsize 
\vspace{5mm}

Author D.D. acknowledges the facilities of Saha Institute of Nuclear Physics, 
Kolkata, India. Author L.S.K. acknowledges support from the P25 group at Los 
Alamos National laboratory.

\end{document}